# Pressure-induced phonon-freezing in the ZnBeSe alloy: a study via the percolation 'mesoscope'

**Gopal K. Pradhan[1], Chandrabhas Narayana[1], Olivier Pagès [2,*], Abedalhasan Breidi[2], Jihane Souhabi[2], Andrei Viktor Postnikov[2], Fouad El Haj Hassan[3], Sudip K. Deb[4], Franciszek Firszt[5], Wojtek Paszkowicz[6] and Abhay Shukla[7]**

[1] *Light Scattering Laboratory, Jawaharlal Nehru Centre for Advanced Scientific Research (JNCASR), Bangalore 560064, India*

[2] *Laboratoire de Physique des Milieux Denses – IJB, Université Paul Verlaine, Metz 57078, France*

[3] *Laboratoire de Physique des Matériaux, Université libanaise, El Hadath, Beyrouth, Liban*

[4] *Indus Synchrotron Utilization Division, Centre for Advanced Technology, Indore 452013, India*

[5] *Institute of Physics, N. Copernicus University, 87-100 Toruń, Poland*

[6] *Institute of Physics, Polish Academy of Sciences, 02668 Warsaw, Poland*

[7] *Université Pierre et Marie Curie-Paris6, CNRS-UMR7590, IMPMC, 140 rue de Lourmel, Paris, F-75015 France*

## Abstract

We use the 1-bond→2-phonon percolation doublet of zincblende alloys as a 'mesoscope' for an unusual insight into their phonon behavior under pressure. We focus on (Zn,Be)Se and show by Raman scattering that the original Be-Se doublet at ambient pressure, of the stretching-bending type, turns into a pure-bending singlet at the approach of the high-pressure ZnSe-like rocksalt phase, an unnatural one for the Be-Se bonds. The 'freezing' of the Be-Se stretching mode is discussed within the scope of the percolation model (mesoscopic scale), with *ab initio* calculations in support (microscopic scale).

---

* Corresponding author, e-mail: pages@univ-metz.fr

Recently, a unified understanding of the phonon behavior of usual $A_{1-x}B_xC$ semiconductor alloys with fourfold-coordinated cubic zincblende (ZB) structure, as observed by Raman scattering and infrared absorption, could be achieved within the so-called percolation model developed by one of us (O. P.). In this model a random alloy is viewed as a composite of the AC- and BC-like regions, where A-C and B-C bonds are highly self-connected, respectively, each region providing one phonon per bond (1-bond→2-mode) [1]. This reveals that phonons give a natural insight into the alloy disorder at the unusual mesoscopic scale. We introduced a terminology that the percolation doublet (with splitting denoted by Δ) acts as a 'mesoscope'. In principle, the change in frequency and intensity of each 1-bond→2-mode mesoscope (AC- and BC-like) may be studied under the influence of any stimulus. This opens the way for a basic understanding beyond that given by the usual 1-bond→1-mode 'macroscope' of Chang and Mitra [2], based on a description of an alloy as a uniform continuum according to the virtual crystal approximation (VCA).

In this work, we explore what the mesoscope may say about the phonon behavior of semiconductor alloys under pressure. The aim is to decide whether the AB- and AC-like regions are affected in the same way by hydrostatic pressure, or whether one region behaves in a specific manner. The most suitable alloy to address such issue is $Zn_{1-x}Be_xSe$. While ZnSe adopts the usual sixfold-coordinated cubic NaCl (metallic rocksalt – RS) phase under pressure, BeSe is one of the few exceptional systems that transforms to the sixfold-coordinated hexagonal NiAs phase [3,4]. Furthermore, the transition pressure is much higher for BeSe (~56 GPa) than for ZnSe (~13 GPa). An interesting question then is what happens to the Be-Se bonds of ZnBeSe crystals with moderate Be content when the ZnSe-like medium begins its natural ZB→RS transition? This, we investigate by taking advantage of the well-resolved Be-Se mesoscope (Δ~40 cm$^{-1}$). We perform Raman measurements at increasing pressure up to the ZB→RS transition using a series of alloys with moderate Be content, and attempt to achieve consistent understanding of the observed effects at both the mesoscopic (percolation scheme) and microscopic (*ab initio* calculations) scales.

$Zn_{1-x}Be_xSe$ single crystals with x values of 0.11, 0.16, 0.24 and 0.55 (the achievable limit so far), measured by x-ray diffraction, were grown by the Bridgman method. For each sample a (111)-oriented cleaved piece was polished to a ~25 μm thick platelet and placed together with ruby chips into a 200 μm thick stainless steel gasket (preindented to 65 μm) inserted between the diamonds of a Mao/Bell-type diamond anvil cell. Methanol/ethanol/water (16:3:1) was used as pressure transmitting medium, the pressure being determined via the ruby fluorescence linear scale. The pressure environment is assured to be hydrostatic till ~14 GPa, beyond which it can be treated quasihydrostatic till ~30 GPa.

Unpolarized Raman spectra were recorded in backscattering geometry on the (111) crystal face with the 532.0 nm excitation. In such geometry both the transverse optical (TO) and longitudinal optical (LO) modes are Raman-active. In the discussion we focus on TO modes, because such mechanical vibrations give reliable insight into the individual percolation-type oscillators present in the crystal [1]. In the labelling of TO modes, the subscript and superscript refer to the bond species and to the host region, respectively. In contrast, the close LO modes of each percolation doublet couple via their macroscopic electric field, resulting in a unique LO mode [1]. The percolation doublet thus disappears in the LO symmetry, and with it the mesoscopic insight.

*Ab initio* calculations of the TO density of states at the Brillouin zone-centre (ZC TO-DOS, which compares to Raman spectra) were done with the SIESTA code. We used a 64-atom supercell containing the 'ultimate' percolation motif (referred to as the 2-imp. motif), i.e. a pair of next-nearest-neighbor impurities (say A) immersed in the environment of the other species (BC-like). Both the ZnBeSe and GaAsP alloys were considered (the latter for reference purpose), using the basis of pseudopotentials set up in Refs [5] and [1], respectively.

For the presentation and discussion of the data we focus on $x=0.24$, but the effects are general. Representative pressure-dependent Raman spectra are shown in Fig. 1. The frequency versus pressure variations of the main features are shown in the inset. A 2TO+1LO percolation signal shows up clearly in the Be-Se spectral range (~550 cm$^{-1}$). In the Zn-Se spectral range (~225 cm$^{-1}$) the assignment is more difficult due to the existence of several modes. Although a preliminary 1TO+1LO assignment was proposed, the current high pressure data lead to a more refined one (see below).

With increasing pressure the intensity of the LO modes decreases, a well-known effect of the progressive metallization of the sample resulting in the ZB→RS transition. The total LO-extinction occurs at ~23 GPa, coinciding with the actual ZB→RS transition observed by x-ray diffraction (not shown). More fascinating effects relate to TO modes as marked by capitals, i.e. the splitting of the lower Zn-Se branch (A), the transient emergence of a weak mode in-between two main Zn-Se branches (B), and the convergence of the lower Be-Se TO branch onto the upper one (C).

First, we focus on the Zn-Se spectral range. A natural reference is the pressure dependence of the Raman frequencies in pure ZnSe, as reproduced in Fig. 2 (symbols, digitalized from Ref. [6]). The two sub-branches that proceed from the 'nominal' TO mode at ~2 and ~7 GPa are usually interpreted as TO modes due to intermediate phases in-between the ZB and RS ones. However, no such phases could be identified in ZnSe yet, neither experimentally nor by *ab initio* calculations [7]. Our view is rather that the entire ZnSe spectrum results from an anharmonic decay of the discrete (zone-centre) TO into a variety of two-phonon continua, at least two (corresponding to the two sub-branches). Characteristic features are discussed below.

In such decay process, conservations of the momentum and energy impose equal and opposite wave vectors of the two phonons, and quasi-resonance of the discrete TO and two-phonon continua. Referring to the phonon DOS of pure ZnSe [8], the likely continua are the 2TA and TA+LA originating from the transverse (TA) and longitudinal (LA) acoustical branches at the K and L Brillouin zone-edges, respectively. Under pressure, the discrete TO and two-phonon continua decouple because different pressure-dependencies lead to a departure from resonance conditions. In ZnSe the zone-edge TA 'softens' under pressure, while the zone-centre TO and zone-edge LA 'hardens' fast [9]. The TA trend reinforces in the 2TA continuum, while it is more or less compensated by the opposite LA trend in the TA+LA continuum. We thus reassign the two branches that proceed from the 'nominal' TO at ~2 and ~7 GPa in pure ZnSe, with large-negative and small-positive pressure coefficients, as the 2TA and TA+LA continua, respectively.

The A and B features in Fig. 1 can then be simply explained by considering that in the alloy a weak extra TO appears and substitutes for the strong (pure-ZnSe like) TO for the anharmonic decay into the 2TA continuum. The principle is outlined in Fig. 2. The weak TO is represented by a thin line downshifted from the strong TO (dashed line) by ~30 cm$^{-1}$. With this, the 2TA-decoupling is delayed from ~2 GPa in pure ZnSe (from the strong TO, see the dotted arrow) to ~7 GPa in the alloy (from the weak TO), in reference to A, while B at ~7 GPa is due to transient appearance of the TA+LA continuum just after decoupling from the strong TO and before re-coupling with the weak TO (see the oval). In fact the coupling-related antiresonance in-between the two TO (solid squares in Fig. 1) disappears in this intermediary decoupling regime (see the spectrum at ~9 GPa in Fig. 1).

Altogether this implies a well-resolved Zn-Se percolation doublet (the weak plus strong TO) in ZnBeSe, instead of a singlet as earlier presumed. The splitting of the Zn-Se doublet (Δ~30 cm$^{-1}$, refer to the inset in Fig. 1), almost as large as the Be-Se one, is disconcerting though [1]. For example in ZnBeTe the Zn-Te (Δ~6 cm$^{-1}$) and Be-Te (Δ~35 cm$^{-1}$) splittings scale in the ratio 1:6. For independent insight we performed *ab initio* calculations with the 2-imp. (Be) motif in ZnSe. The short Be-Se bonds create a local tension in ZnSe, giving rise to a local Zn-Se mode around Be ($TO_{Zn-Se}^{Be}$, weak – see the vibration pattern in Fig. 2) at a lower frequency than that of bulk ZnSe ($TO_{Zn-Se}^{Zn}$, strong). As expected, the *ab initio* splitting is small, of ~8 cm$^{-1}$, corresponding to a position of the harmonic (uncoupled) weak $TO_{Zn-Se}^{Be}$ renormalized to the thick line in Fig. 2. We infer that the anharmonic coupling induces a massive frequency downshift of the weak $TO_{Zn-Se}^{Be}$ with respect to the *ab initio* prediction (by ~22 cm$^{-1}$), the strong $TO_{Zn-Se}^{Zn}$ remaining unaffected. This suggests that the continua couple strongly to the weak TO and weakly to the strong one.

For a quantitative insight, we perform analysis of the TO lineshapes using the formula

$I(\omega) \sim [\alpha+\Gamma(\omega)]^2 \cdot [(\omega-\omega_0-\Delta(\omega))^2+(\alpha+\Gamma(\omega))^2]^{-1}$ [10], where $\omega_0$ and $\alpha$ are the frequency and natural damping of the harmonic TO, and $\Delta(\omega)$ and $\Gamma(\omega)$ relate to the two-phonon DOS and to its Kramers-Krönig's transform, respectively. The two-phonon DOS is designed as a semi-ellipse, its position, magnitude and spectral extension being fixed by its upper end $\omega_c$, and its semi-axes D and a, respectively. The model includes a fourth-order coupling $V_4$ (between pairs of phonons in the two-phonon continuum) on top of the third-order one $V_3$ (between the TO and the two-phonon continuum). To make it physically more meaningful only one parameter is adjusted per TO mode, i.e. the strength $V_3^2 D$ of the cubic coupling, the criterion for validation being the ability to reproduce the TO-decoupling process (at P~7 GPa for both TO modes).

For the strong $TO_{Zn-Se}^{Zn}$, $\omega_0$ is taken as the observed peak position, since anharmonic effects are presumably small, while for the weak $TO_{Zn-Se}^{Be}$ we use the *ab initio* value. As for the two-phonon DOS, the pressure dependence of the upper ends ($\omega_c$) are derived by linear interpolations of the pressure-dependencies of the decoupled 2TA (for $TO_{Zn-Se}^{Be}$) and LA+TA (for $TO_{Zn-Se}^{Zn}$) bands in pure ZnSe (refer to the dotted lines converging to ~220 cm$^{-1}$ at 0 GPa in Fig. 2), after slight translations (by less than 2 cm$^{-1}$) needed to mimic alloying effects. The lower end, monitored by a once $\omega_c$ is known, provides an antiresonance on the low-frequency tail of the coupled TO (coupling regime) or uncoupled-two-phonon-DOS (decoupling regime). The 2TA-antiresonance is clearly visible in our Raman spectra (refer to open squares in Fig. 1). For an insight into the (TA+LA)-related antiresonance we resort to the pure-TO Raman spectrum (14%Be, 0 GPa) available in Ref. [11] (top curve of Fig. 2). Clear antiresonances on each side of the strong TO therein, spaced by ~15 cm$^{-1}$, correspond to a~5 cm$^{-1}$, taken constant for all pressures. Last, the strength $V_4 D$ of the fourth-order coupling is constrained to the $\alpha$ value (within 5%), taken as ~2.5 cm$^{-1}$ for all pressures.

As expected, $V_3^2 D$ is larger for the weak $TO_{Zn-Se}^{Be}$ (with a value of ~90 cm$^{-1}$) than for the strong $TO_{Zn-Se}^{Zn}$ (~10 cm$^{-1}$). The pressure dependent line shapes of the weak TO (the strongly coupled one) over its decoupling process are shown in the inset of Fig. 2. The main features are well-reproduced, i.e. the deepening of the antiresonance at ~180 cm$^{-1}$ just when decoupling occurs (refer to open squares in Figs 1 and 2 – the arrow marks the frequency of the harmonic TO therein, for reference purpose), and the consequent appearance of the uncoupled 2TA band as an intense peak (refer to the star in Fig 1 and 2) before it collapses into a damped feature away from the TO mode.

In brief, an anharmonic coupling greatly magnifies (by ~4) the natural splitting ($\Delta$~8 cm$^{-1}$) of the Zn-Se percolation doublet, giving a chance to observe its dependence on pressure. The two

ZnSe-like TO branches are quasi-parallel in Fig. 1 (spaced by ~30 cm$^{-1}$, refer to the inset), indicating a constant Zn-Se splitting. *Ab initio* calculations at 0 and 10 GPa with the 2-imp. (Be) motif in ZnSe confirm the trend (not shown). We conclude that, as far as the Zn-Se bonds are concerned, the BeSe- and ZnSe-like regions of the crystal behave similarly under pressure.

However, this is not the case for the Be-Se bonds, as testified by singularity C. We emphasize that C is not composition-dependent, i.e. it occurs at the same critical pressure of 14±1 GPa for all our samples, indicating a local origin at the bond scale. In fact, the ZC TO-DOS per Be atom obtained with the 2-imp. (Be) motif in ZnSe at 0 and 10 GPa do exhibit singularity C. The curves are shown in the body of Fig. 3, specifying the vibration pattern per mode. In view of these, we identify the microscopic mechanism behind C as the transformation of $TO_{Be-Se}^{Be}$ (vibration along the Be-Se-Be chain, ∥ chain) from the bond-stretching type to the bond-bending type under pressure. The nature of the doubly-degenerate $TO_{Be-Se}^{Zn}$ (vibrations perpendicular to the Be-Se-Be chain, in-plane and out-of-plane, ⊥ chain), as for it, remains unchanged, of the bond-bending type.

For more insight we performed similar *ab initio* calculations with the 2-imp. (Zn) motif in BeSe (Zn-dilute limit, not shown). The long Zn-Se bonds create a local compression in BeSe, leading to a local Be-Se mode close to Zn, i.e. $TO_{Be-Se}^{Zn}$ (the vibration pattern is that shown in Fig. 2, where Zn and Be are exchanged), at a higher frequency than the bulk BeSe mode, i.e. $TO_{Be-Se}^{Be}$. This gives a clear Be-Se percolation doublet, though the splitting is less than in the Be-dilute limit, of ~11 cm$^{-1}$. Now, we observed the same splitting at 0 and 10 GPa, indicating that singularity C disappears when the Be-Se bonds are dominant in the alloy, thus not influenced by the Zn-Se bonds.

A further instructive case is that of GaAsP. Its short bond (Ga-P) exhibits a distinct percolation doublet (Δ~12 cm$^{-1}$) [1], as in ZnBeSe, but, the parents (GaAs, GaP) adopt the same phase under pressure (ZB→Cmcm), and moreover at nearly the same critical pressure (~15±3 GPa). The ZC TO-DOS per P atom obtained at 0 and 10 GPa with the 2-imp. (P) motif in GaAs are shown in the inset of Fig. 3. The Ga-P percolation doublet is shifted upwards as a whole when approaching the phase transition, and no C-type convergence is observed.

It emerges that singularity C is not intrinsic to short bonds in alloys but specific to the Be-Se bonds in ZnBeSe when their fraction is such (low or moderate) that they are forced to adopt the ZnSe-like RS phase under pressure.

For an intuitive discussion of the physics behind C, we turn to the percolation model. Note that such phenomenological model relies on a scalar description of the alloy (linear chain approximation), so that everything comes down to a question of bond-stretching forces only, by construction. Still, a parallel between the percolation model (mesoscopic scale) and *ab initio* calculations (microscopic scale) can be drawn by realizing that *'bond-bending within a given* *impurity motif corresponds to bond- stretching of the (like) surrounding bonds from the host matrix, and vice versa'*. If we refer to the 2-imp. (Be) motif in ZnSe, this comes down to discuss the low- and high-frequency Be-Se modes in terms of Be-Se stretching within the BeSe- and ZnSe-like regions, respectively (see the shaded areas in the vibration patterns, Fig. 3), consistently with the terminology of the percolation model. Within such stretching-type model, feature C can thus be understood as due to the progressive 'freezing' of the Be-Se bonds from the minor BeSe-like region when this is forced to adopt the unnatural RS phase of the host ZnSe-like region. The oscillator strength is transferred to the close Be-Se bonds of the latter region.

In brief, the percolation mesoscope reveals that the lattice dynamics of the ZnBeSe crystal basically changes when approaching the ZB→RS ZnSe-like transition under pressure, an unnatural one for the Be-Se bonds. Within the scope of the percolation model, the highly self-connected Be-Se bonds from the BeSe-like region 'freeze', the oscillator strength being channelled to the less self-connected Be-Se bonds of the surrounding ZnSe-like region. This process is transparent for the Zn-Se dynamics, and completed (~14 GPa) before the transition to RS occurs (~23 GPa at 24%Be). Generally this work shows how the percolation mesoscope may help to achieve further understanding of the pressure dependence of phonons in alloys beyond the usual VCA paradigm.

**Figures**

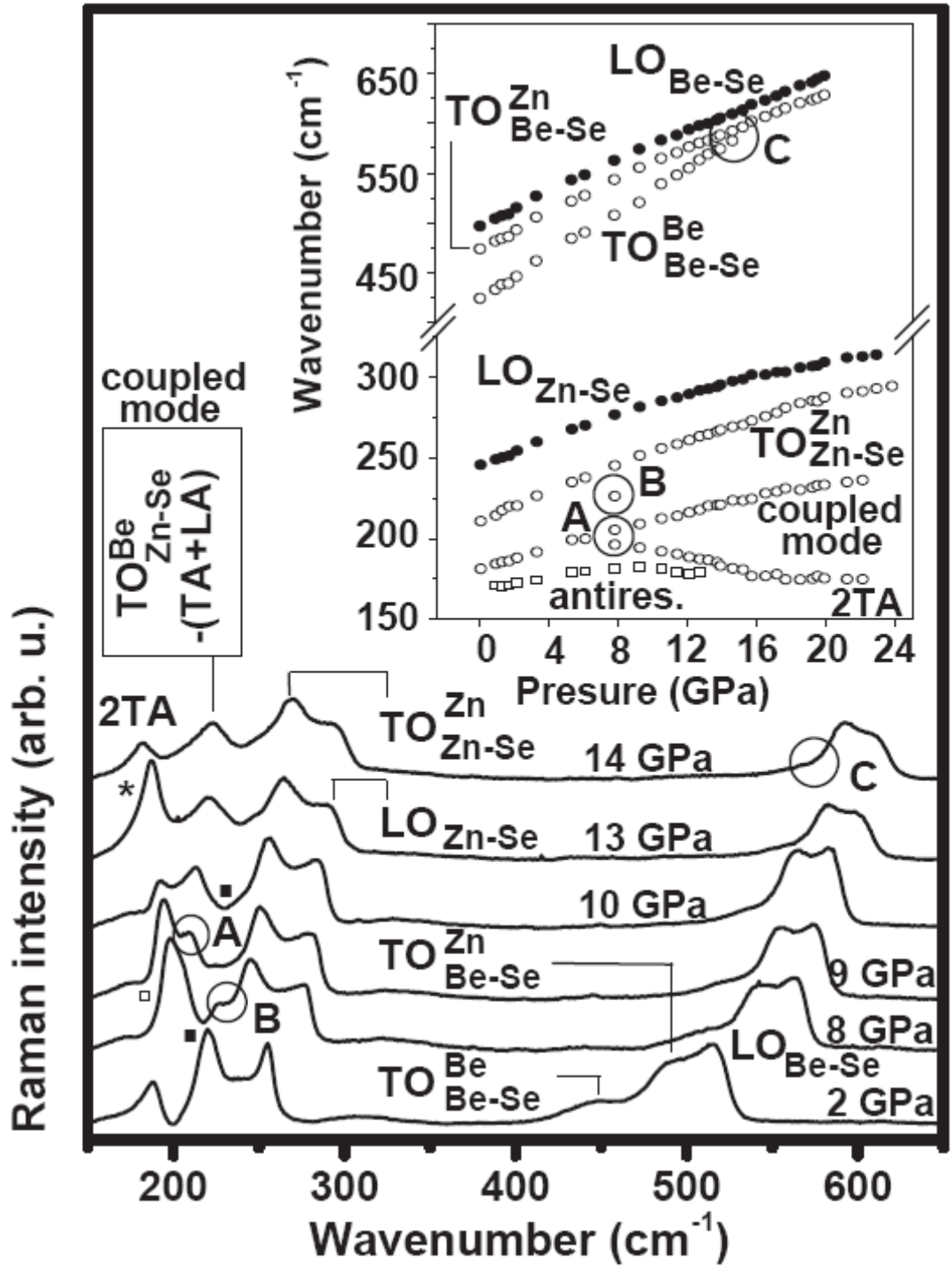

**FIG. 1.** Pressure-dependent Raman spectra of $Zn_{0.76}Be_{0.24}Se$. The frequency versus pressure variations of the main features are shown in the inset. Remarkable trends (A,B,C) are marked by circles. The antiresonances (squares) and the strong peak (star) are characteristic of anharmonic coupling.

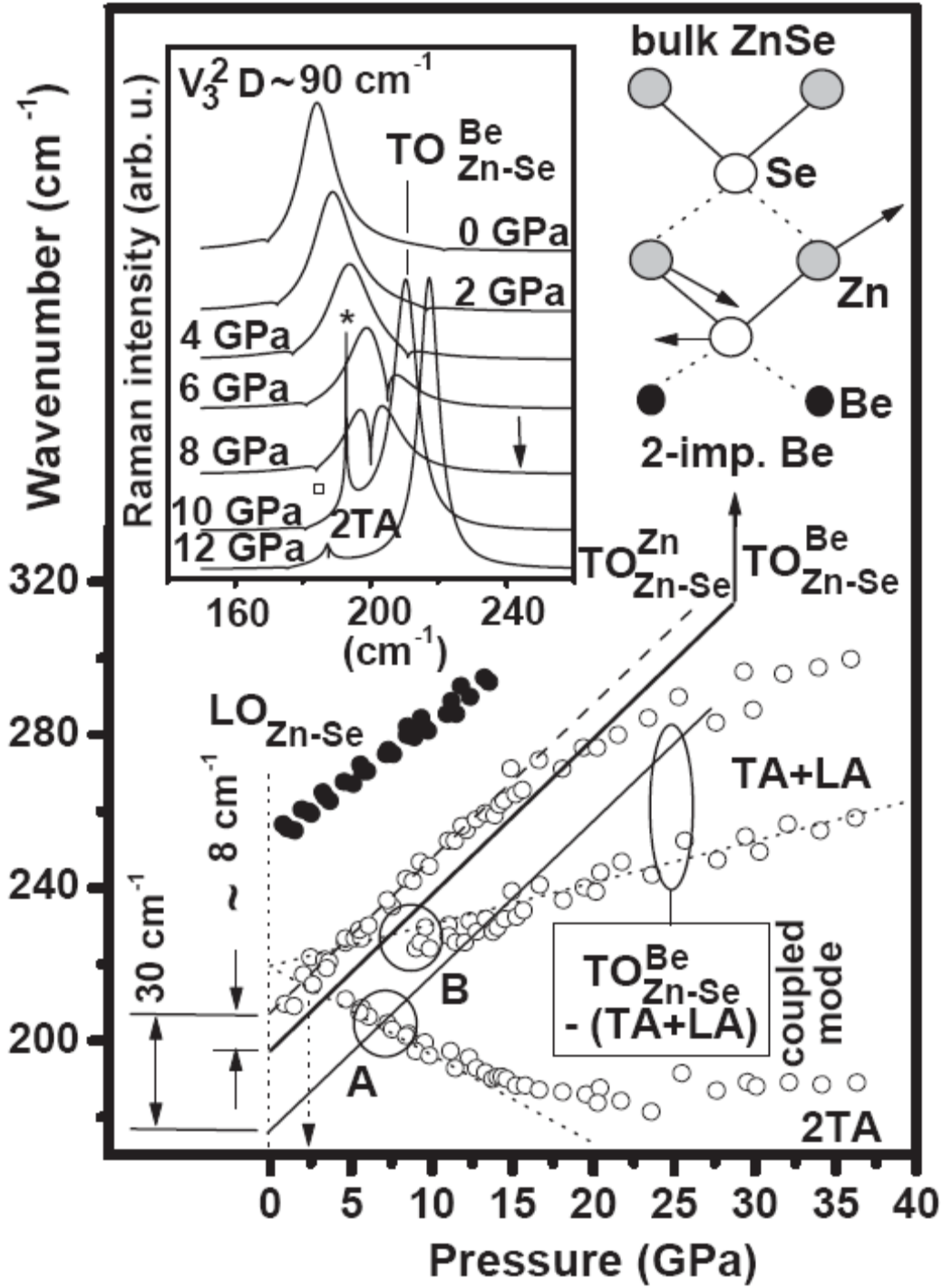

**FIG. 2.** Schematic explanation of features A and B (circles) by referring to pure ZnSe (symbols, digitalized from Ref. [6]). In the alloy the weak $TO^{Be}_{Zn-Se}$ (thick line – in the vibration pattern the dotted and plain lines indicate perpendicular planes) appears on top of the strong $TO^{Zn}_{Zn-Se}$ (pure-ZnSe like, dashed line). The natural splitting of ~8 cm$^{-1}$ is magnified to ~30 cm$^{-1}$ by an anharmonic coupling that downshifts the weak mode from the thick line to the thin one. The lineshapes of the weak mode over its decoupling process are shown in the inset. The star and squares mark features characteristic of anharmonic coupling, as in Fig. 1. At 8 GPa an arrow marks the position of the harmonic mode, for reference purpose.

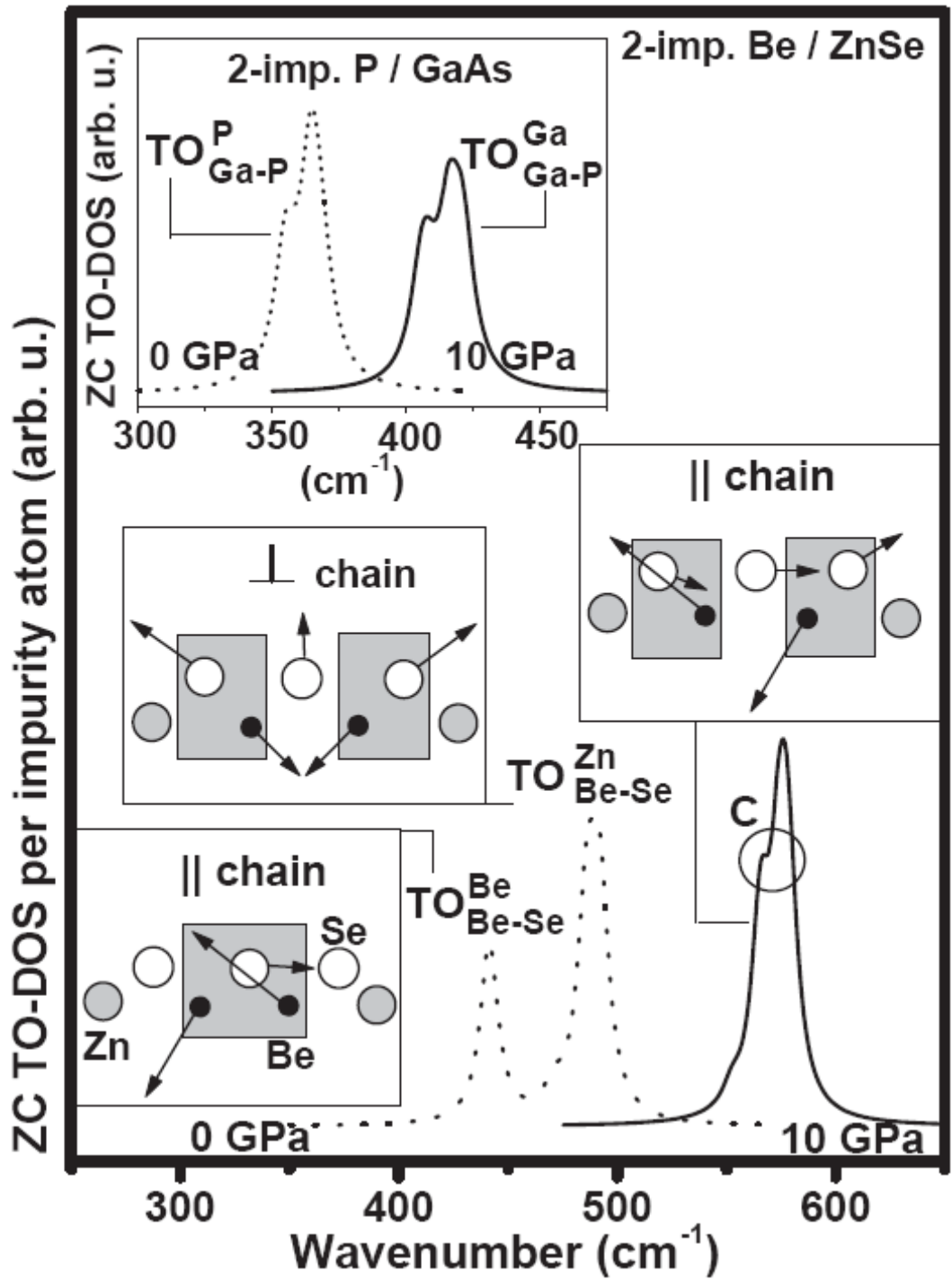

**FIG. 3.** *Ab initio* ZC TO-DOS per impurity of the 2-imp. Be (main curves) and P (inset) motifs in ZnSe- and GaAs-like supercells, respectively. The Be-Se vibration patterns are indicated, emphasizing the bond-stretching modes (dashed areas) in reference to feature C (circle).

11